\documentclass[11pt,xcolor=table]{article}

\usepackage{adjustbox,amssymb,amsmath,amsfonts,array,authblk,booktabs,enumitem,eurosym,geometry,ulem,graphicx,caption,color,natbib,rotating,setspace,sectsty,comment,footmisc,caption,natbib,
pdflscape,placeins,array,hyperref}
\usepackage[flushleft]{threeparttable}

\usepackage{soul}
\usepackage{bbm}
\usepackage[english]{babel}
\usepackage{hhline}
\usepackage{multirow}
\usepackage{scalerel}
\usepackage[table,xcdraw]{xcolor}
\usepackage{geometry}
\usepackage{setspace}
\usepackage{amsbsy}
\usepackage{appendix}
\usepackage{soul}
\usepackage{float}
\usepackage{subcaption}
\usepackage{graphicx}

\setcitestyle{authoryear,open={(},close={)}} %Citation-related commands
\usepackage{lineno}

\providecolor{added}{rgb}{0,0,1}
\providecolor{deleted}{rgb}{1,0,0}

\newcolumntype{L}[1]{>{\raggedright\let\newline\\arraybackslash\hspace{0pt}}m{#1}}
\newcolumntype{C}[1]{>{\centering\let\newline\\arraybackslash\hspace{0pt}}m{#1}}
\newcolumntype{R}[1]{>{\raggedleft\let\newline\\arraybackslash\hspace{0pt}}m{#1}}

\begin{document}

\begin{titlepage}
% \title{The Zero Conditional Mean Assumption in Linear Econometric Models}
% \title{On the Zero Conditional Mean Assumption in Linear Econometric Models}
% \title{Zero conditional mean assumption is not sufficient for causal inference}
\title{On the Role of the Zero Conditional Mean Assumption for Causal Inference in Linear Models\footnote{We thank Christian Dahl, Markus Fr\"{o}lich, Arthur Lewbel, Ye Lu,  Francesca Molinari, Pedro Sant'Anna and Rami Tabri for useful conversations.}}

%\title{What does OLS identify under the zero conditional mean assumption?}
\author[1]{Federico Crudu} \author[2]{Michael C. Knaus} \author[3]{Giovanni Mellace} \author[4]{Joeri Smits}

\affil[1]{University of Siena and CRENoS}
\affil[2]{University of T\"ubingen}
\affil[3]{University of Southern Denmark}
\affil[4]{Harvard University}

\date{\today}
\maketitle
\begin{abstract}
\noindent Many econometrics textbooks imply that under mean independence of the regressors and the error term, the OLS parameters have a causal interpretation. We show that even when this assumption is satisfied, OLS might identify a pseudo-parameter that does not have a causal interpretation. Even assuming that the linear model is “structural” creates some ambiguity in what the regression error represents and whether the OLS estimand is causal. This issue applies equally to linear IV and panel data models. To give these estimands a causal interpretation, one needs to impose assumptions on a “causal” model, e.g., using the potential outcome framework. This highlights that causal inference requires \textit{causal}, and not just \textit{stochastic}, assumptions.
\\
\vspace{1em}

\bigskip
\end{abstract}
\setcounter{page}{0}
%\thispagestyle

%\newcommand\blfootnote[1]{%
  %\begingroup
  %\renewcommand\thefootnote{}\footnote{#1}%
  %\addtocounter{footnote}{-1}%
  %\endgroup
%}
%
%\blfootnote{}
\end{titlepage}
\pagebreak \newpage

\doublespacing

\section{Introduction}\label{sec:intro}

The zero conditional mean assumption imposing that the regression error is mean independent of the covariates and has mean zero is a building block of linear regression models. Under this assumption, many econometrics textbooks give the model coefficients a causal interpretation \citep{angrist2017undergraduate}. For example, the econometrics textbook most used in economics, \citet{wooldridge2019introductory}, discusses the example of yield as outcome and fertilizer as regressor as follows ``if fertilizer amounts are chosen independently of other features of the plots, then [zero conditional mean error] will hold: the average land quality will not depend on the amount of fertilizer. However, if more fertilizer is put on the higher-quality plots of land, then the expected value of $u$ [the error] changes with the level of fertilizer, and [zero conditional mean error] fails.''.\footnote{According to \href{https://opensyllabus.org/result/field?id=Economics}{Opensyllabus.org} accessed on March 18th 2022.} As we will show, the first sentence in this quote is true, but the second not necessarily. Notice that a similar approach is used in other textbooks, e.g., \cite{cameron2005microeconometrics}, \citet{hayashi2000econometrics}, and \citet{stock2019introduction}.

\textcolor{black}{Although the literature has recognized the advantages of the potential outcome framework over the traditional linear modeling approach \citep[see the widely cited article by][]{IW09}, we clarify the source of ambiguity that may stem from using a non causal model. In particular,} this note highlights that there are cases where the zero conditional mean assumption holds, but the parameter estimated via OLS suffers from omitted variable bias and has no causal interpretation. To illustrate this, note that any regression with a fully saturated model would always be causal if the zero conditional mean assumption would be sufficient for the causal interpretation of the parameters. Consider the following two examples:
\begin{itemize}
	\item $wage_i = \beta_0 + \beta_1 college_i + u_i$ where $college_i$ is a binary indicator. In this model, it is always true that $E[u_i|college_i] = 0$, but no economist would give $\beta_1$ a causal interpretation.
	\item $wage_i = \beta_0 + \beta_1 college_i + \beta_2 good\_health_i + \beta_3 college_i \times good\_health_i + u_i$ where $good\_health_i$ is a binary indicator. $E[u_i|college_i,good\_health_i] = 0$  always holds because the model is fully saturated. However, no economist would give $\beta_1$ a causal interpretation.
\end{itemize}

In our opinion, the ambiguity occurs  because the difference between unobserved variables in a causal model and the statistical error in the statistical model is often not made explicit. Many books use the same error term to discuss both statistical properties of OLS as an estimator and omitted variable bias. This might induce in the reader the false idea that zero conditional mean error implies no confounding. However, the reverse is true, that is, no confounding implies zero conditional mean error. While the lessons learned in both estimation and identification are still valid, we think that the conceptual ambiguity is worth clarifying and to be kept in mind when teaching econometrics. 

The goal of this note is to exactly show where and why the ambiguity arises and to propose alternative and less ambiguous ways of presenting the assumptions needed for a causal interpretation of the linear model coefficients. We also show how linear instrumental variable (IV) and panel data models are affected by the same issue. In the Online Appendix we also report several numerical examples to further illustrate our results.  %To this end we first illustrate why taking the purely stochastic zero conditional mean assumption imposed on the \textit{observed outcome} is not sufficient to identify causal effects, but imposing it on the \textit{potential outcome} is. We also discuss alternative ways to motivate estimation of causal effects using a linear model.

\section{Why the zero conditional mean assumption is not sufficient}\label{sec:ols}

Consider the simple regression
\begin{equation}
Y_i=\gamma+\lambda D_i+\varepsilon_i \label{reg}
\end{equation}

and assume zero conditional mean of the error term, i.e., $E[\varepsilon_i|D_i]=0$. From a purely statistical point of view this is equivalent to assuming that the conditional expectation of $Y$ given $D$ is linear, i.e. $E[Y_i|D_i]=\gamma +\lambda D_i$. We now show that this is not sufficient for a causal interpretation of $\lambda$, which can be arbitrarily different from the causal effect of $D$ on $Y$. To this end, let $Y_i(d)$ be the potential outcome of individual $i$ if $D_i=d$ and assume that 
$$
Y_i(d)=\alpha+\tau d+ \beta U_i.
$$
This implies that the causal effect of a one-unit increase in $D$ on $Y$ is constant and equal to $\tau$, i.e. $Y_i(d+1) - Y_i(d)=\tau$. 
%To clarify the role of $D$, $U$, and $Y$ consider the following DAG:
%
%\begin{figure}[h!]{
%\centering
%\includegraphics[scale=0.5]{DAG2-1.pdf}
%\caption{Confounding path and direct causal path with no additional variables.}\label{fig:dag1}
%}
%\end{figure}
%The first is the direct effect $D\to Y$, which is what we want to identify. The second is the fork structure $Y \leftarrow U \to D$, which is a confounding path.

For example, $Y_i(d)$ could represent the wage individual $i$ would receive if she had completed $d$ years of education and $U_i$ is unobserved ability.
Assume that $E[U_i|D_i]=\mu+\delta D_i$  such that we can write $U_i=E[U_i|D_i]+\nu_i=\mu+\delta D_i+\nu_i$ with $E[\nu_i|D_i]=0$. Notice that this assumption is always satisfied if $D$ is binary or if $U$ and $D$ are drawn from a bivariate normal distribution. Under the stable unit treatment value assumption (SUTVA), i.e. $Y_i=Y_i(d)$ if $D_i=d$, we have:
\begin{eqnarray*}
Y_i=Y(D_i)&=&\alpha+\tau D_i+ \beta [\mu+\delta D_i + \nu_i],\\
&=&\underbrace{\alpha+ \beta\mu}_{\gamma} +\underbrace{(\tau+\beta \delta )}_{\lambda}D_i+\underbrace{\beta \nu_i}_{\varepsilon_i}.
\end{eqnarray*}
Therefore,  the conditional expectation of the observed outcome given $D$ is linear, $E[Y_i|D_i]=\gamma +\lambda D_i$, and the zero conditional mean assumption, $E[\varepsilon_i|D_i]=\beta E[\nu_i|D_i]=0$, is satisfied.
However, although $\lambda=\tau+\beta \delta$ represents how a change in $D$ changes $E[Y|D]$, it can be arbitrarily different from the causal effect of $D$ on $Y$, which is given by $\tau$. 
Notice that $\beta \delta$ is the usual omitted variable bias that is well known and discussed in any econometric textbook. Most books imply that if $D$ is uncorrelated with $\varepsilon$ there is no omitted variable bias. This is contradicted by our derivation showing that $E[\varepsilon_i|D_i]=0$, and as a consequence $Cov[\varepsilon_i,D_i]=0$, can hold in the presence of omitted variable bias.  

We conjecture that this tension is due to the notion that any omitted variable would be included in the error term and that this would violate the zero conditional mean assumption (see the quote in the introduction). However, it is often not clear which error term is referred to. There is the statistical error term $\varepsilon$ and the omitted variable in the causal model $U$, but they are not necessarily the same, i.e.
$$
\varepsilon_i\equiv Y_i-E[Y_i|D_i] = \beta \nu_i \neq \beta U_i
$$

in our example. This ambiguity disappears only in the absence of the omitted variable $U$, i.e., the statistical error is the only unobserved part of the potential outcome. In that case, assuming zero conditional mean error suffices for causal interpretation. 

Notice that $E[\varepsilon_i|D_i]=0$ does not imply that the conditional mean independence assumption on the potential outcome holds, i.e $E[Y(d)|D_i=d]=E[Y(d)|D_i=d'], d\neq d'$. Indeed, $E[Y(d)|D_i=d]-E[Y(d)|D_i=d']=\beta \delta(d-d')$. Therefore this assumption on the potential outcomes is only satisfied if there is no omitted variable bias, so that $\beta \delta =0$. However, it \textit{would} be satisfied if we put a zero conditional mean assumption on the unobservable part of the potential outcome, i.e. $E[\beta U_i|D_i] = 0$. This shows that assuming zero conditional mean of the unobserved part in the \textit{potential outcome} is sufficient for a causal interpretation. 

To avoid such confusion, it is important to emphasize that exogeneity cannot be used in a mere statistical model but needs to refer to a causal model instead. In Section \ref{sec:alt} we propose some alternative formulation of the assumption needed that do not necessarily require introducing the potential outcome framework.

Adding control variables does not change our results as we show in the \hyperref[app:cov]{Appendix} as well as in the simulations we run in the Online Appendix.

\section{Implications for linear IV and panel data models}\label{sec:iv}
The consequences for linear IV models of our results are twofold. First, the way standard IV is presented in most books and scientific articles is as a solution to the so called ``endogeneity'' issue, i.e., correlation between (some of) the regressor(s) - $D$ in our example of equation (\ref{reg}), and the regression error, $\varepsilon$ in equation (\ref{reg}).
As our example demonstrates, even if $D$ is ``exogenous'', i.e., uncorrelated with  $\varepsilon$, we cannot exclude the presence of omitted variables that make  $\lambda$ a pseudo-parameter that does not have a causal interpretation.  
Second, consider again the regression model 
\begin{equation*}
Y_i=\gamma+\lambda D_i+\varepsilon_i, 
\end{equation*}

even if $D$ is endogenous i.e.,  $Cov[D,\varepsilon]\neq 0$,   estimating $\lambda$ by 2SLS using an instrument $Z$ that is uncorrelated with $\varepsilon$ does not guarantee a causal interpretation. 
To see this, assume that  $Y_i(d)=\alpha+\tau d+ \beta U_i$, $D_i=\pi Z_i + \eta_i$, and $U_i=\mu+\delta Z_i + \xi_i$ with $E[\eta_i|Z_i]=E[\xi_i|Z_i]=0$ such that $E[D_i|Z_i]=\pi Z_i$, and $E[U_i|Z_i]=\mu+\delta Z_i$. Following a similar derivation as in Section \ref{sec:ols}, we can write the observed outcome as 
\begin{eqnarray*}
Y_i=Y(D_i)&=&\alpha+\tau (\pi Z_i + \eta_i)+ \beta (\mu+\delta Z_i + \xi_i),\\
% &=&\alpha+\tau \pi Z_i+ \beta [\mu+\delta Z_i]+ \underbrace{\beta \nu_i+ \pi\eta_i}_{\varepsilon_i},\\
&=&\underbrace{\alpha+ \beta\mu}_{\gamma} +\underbrace{\left(\tau+\frac{\beta \delta}{\pi} \right)}_{\lambda}\pi Z_i+\underbrace{\tau\eta_i + \beta \xi_i}_{\varepsilon_i}.
\end{eqnarray*}

This implies that
\begin{eqnarray*}
E[Y_i|Z_i] &=&\underbrace{\alpha+ \beta\mu}_{\gamma} +\underbrace{\left(\tau+\frac{\beta \delta}{\pi} \right)}_{\lambda}\pi Z_i.
\end{eqnarray*}
Therefore, $E[\varepsilon_i|Z_i]=0$ is satisfied, but $\lambda$, the coefficient of a 2SLS regression of $Y$ on $D$ using $Z$ as an instrument, does not have a causal interpretation unless $\beta \delta =0$. \citet{hahn2011conditional} provide an example of a data generating process (DGP) involving a non-linear (and non-separable) outcome equation, whereby 2SLS identifies a pseudo-parameter that has no causal interpretation. Our derivation shows that the DGP does not need to feature a non-linearity or non-additivity for the 2SLS estimand to lack a causal interpretation despite $E[\varepsilon_i|Z_i]=0$. See the Online Appendix for a numerical illustration of these results.

In panel data models it is often implied that in the regression
\begin{equation}
Y_{it}=\gamma_i+\lambda D_{it}+\varepsilon_{it}, \label{reg3}
\end{equation}
assuming $E[\varepsilon_{it}| D_{it},\gamma_i]=0$ is sufficient to give  $\lambda$ a causal interpretation.
      
Without loss of generality, assume that we only have two periods ($t=1,2$) and that  
$$
Y_{it}(d)=\alpha_i+\tau d+ \beta U_{it}.
$$
Letting $\Delta$  be the first difference operator such that $\Delta W_i= W_{i2}-W_{i1}$, we have that first difference in the potential outcome for a given value of $\Delta D_i$, $d'$, is given by
$$
\Delta Y_i(d')=\tau d'+ \beta \Delta U_i.
$$
Assume that $E[\Delta U_i|\Delta D_i]=\mu+\delta \Delta  D_i$.
Thus, following a similar derivation as in Section \ref{sec:ols}, we have that
\begin{eqnarray*}
E[\Delta Y_i|\Delta D_i] &=&\underbrace{\beta\mu}_{\gamma} +\underbrace{(\tau+\beta \delta )}_{\lambda} \Delta D_i.
\end{eqnarray*}
This shows that $E[\varepsilon_{it}| D_{it},\gamma_i]=0$ is satisfied but $\lambda$ does not have a causal interpretation unless $\beta \delta =0$. 
\section{Alternatives to the zero conditional mean assumption}\label{sec:alt}
Many econometrics textbooks, especially at the undergraduate level directly define $\varepsilon_i\equiv Y_i-E[Y_i|D_i]$. Other books like  \citet{wooldridge2010econometric} define the regression error as the unobserved part of the outcome, $\beta U_i$ in our example, but implicitly rely on the assumption that $\varepsilon_i=\beta U_i$, which we have proven not to be  always satisfied. 
This shows  that it is not enough to assume that the linear regression model is structural to solve the potential ambiguity. Consider again the potential outcome of Section \ref{sec:ols}: 
$Y_i(d)=\alpha+\tau d+ \beta U_i$. As we showed, the observed outcome can be written in two alternative ways: either as $Y_i=\alpha+\tau D_i+ \beta U_i$, or as $Y_i=\gamma+\lambda D_i+\varepsilon_i$. Both equations represent the same structural model. Thus, when imposing a conditional zero mean assumption one needs to be very careful in explaining what is meant by the unobservable part. In fact, both  $\beta U_i$  and $\varepsilon_i$ are possible structural errors depending on how we write the model. 
 
Therefore, if one wants to avoid introducing a causal framework, such as the potential outcomes, it has to be clear that the structural error is meant to include all unobserved variables including omitted variables that might affect both $D$ and $Y$. Our derivation also shows the importance of invoking SUTVA. If this assumption fails, even if we carefully distinguish between statistical and structural errors, we would not be able to interpret the model coefficients as causal. We suggest introducing this assumption early on in the definition of the problem.

Another possibility is to make clear that a marginal effect is not necessarily \textit{causal} and that for causal inference one needs to have a \textit{causal} model, and move the discussion around the identification of causal effects to where causal models are actually introduced.

\section{Conclusions}
We have shown that having exogenous (mean independent) regressors in a linear regression model it is not sufficient to avoid omitted variable bias. Although OLS will consistently estimate the true regression coefficient, the latter does not in general have a causal interpretation. This has strong implications for the way we teach econometrics and extends also to linear IV and panel data models, where it is even more common to rely on a lack of correlation or mean independence between the instrument/regressors and the error term of the observed outcome equation, to make causal claims. Ultimately, our results demonstrate that causal inference requires \textit{causal} rather than merely \textit{stochastic} assumptions. 

\singlespacing
\setlength\bibsep{0pt}
\bibliography{References}

\begin{appendices}
\section*{Appendix: Adding covariates}\label{app:cov}
Consider the regression
\begin{equation*}
Y_i=\gamma+\lambda D_i+\varphi X_i+\varepsilon_i, %\label{reg2}
\end{equation*}
and assume $E[\varepsilon_i|D_i,X_i]=0$.

Assume further that,
\begin{eqnarray*}
	Y_i(d)&=&\alpha+\tau d+ \phi X_i+\beta U_i,\\
	E[U_i|D_i, X_i]&=&\mu+\delta D_i+\pi X_i.  
\end{eqnarray*}

Using a similar derivation as above, we have
\begin{eqnarray*}
	E[Y_i|D_i,X_i] &=&\underbrace{\alpha+ \beta\mu}_{\gamma} +\underbrace{(\tau+\beta) \delta }_{\lambda}D_i+\underbrace{(\phi+\beta \pi )}_{\varphi}X_i
\end{eqnarray*}
	
Once again, despite the fact that $E[\varepsilon_i|D_i,X_i]=0$ is satisfied, $\lambda$ does not have a causal interpretation unless $\beta \delta =0$.  
	
\end{appendices}

\newpage
\begin{appendices}
\counterwithin{figure}{section}
\counterwithin{table}{section}

\huge \noindent \textbf{Online Appendices}

\normalsize
\doublespacing

\section{Introduction}\label{sec:intro}
In this document we study the behavior of the least squares (either ordinary or two-stage) estimator under the data generating process (DGP) described in Section 2 of the main text. In addition, we investigate the case where control variables are included (Appendix to the main text). To clarify the role played by the variables in the simulation experiments we accompany the DGPs with their corresponding directed acyclic graph (DAG) representation. 

\section{The DAG point of view} \label{sec:DAG}
\begin{figure}[ht]{
\centering
\includegraphics[width=0.6\columnwidth]{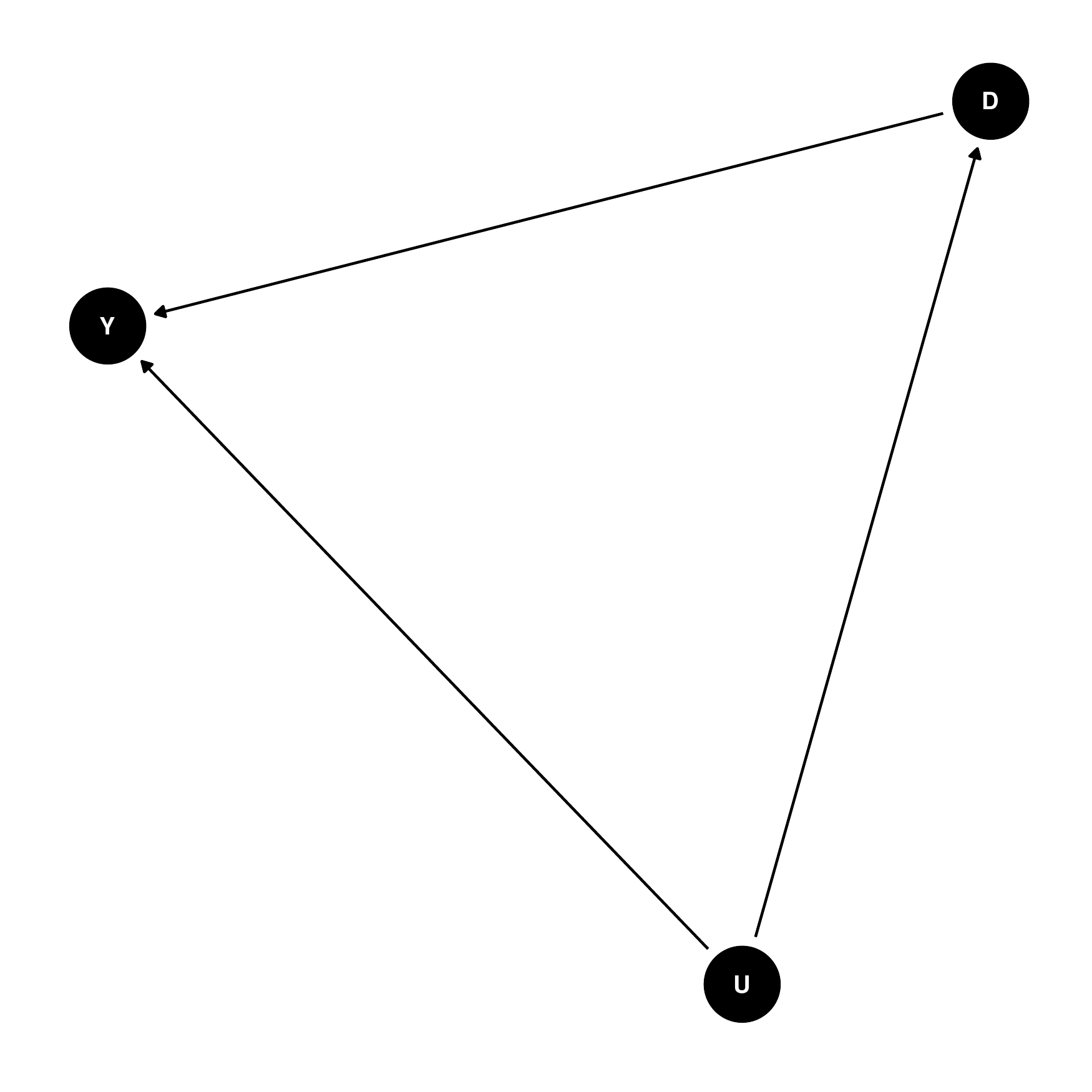}
\caption{Confounding path and direct causal path with no additional variables.}\label{fig:dag1}
}
\end{figure}

The DAG representation of the models presented in the main text (Figure \ref{fig:dag1}) features two elementary relationships.
The first is the direct effect $D\to Y$, which is what we want to identify. The second is the fork structure $Y \leftarrow U \to D$, which is a confounding path. By adding the (observable) variable $X$ we may obtain different graphical structures; either way by controlling for $X$ we open no further confounding paths. Without controlling for the common cause $U$, though, the identification of the direct effect is impossible.

\begin{figure}
\centering
\begin{subfigure}{0.49\textwidth}
\centering
\includegraphics[width = \textwidth]{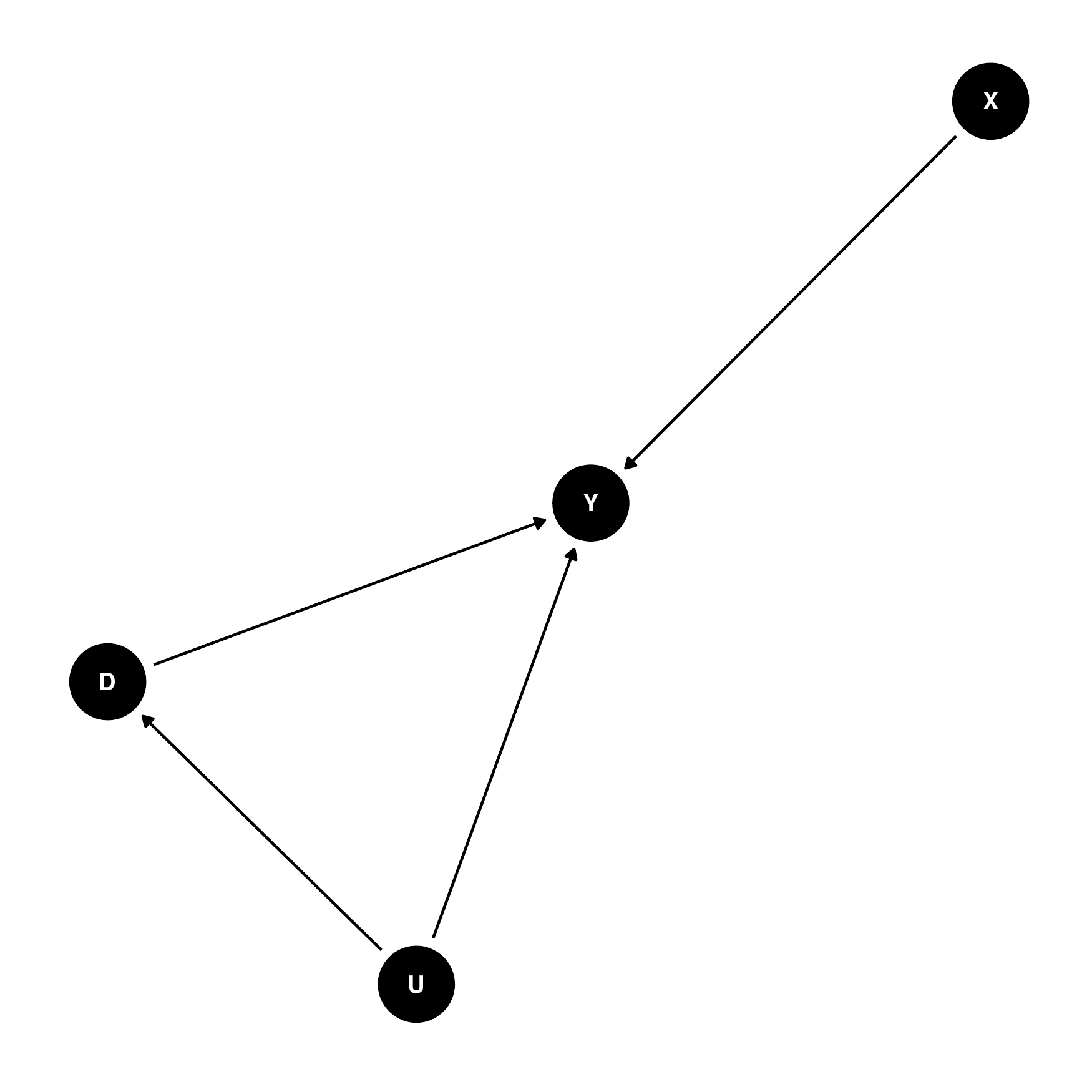}
\caption{$X$ only affects Y.}
\label{fig:dag2}
\end{subfigure}
\begin{subfigure}{0.49\textwidth}
\centering
\includegraphics[width = \textwidth]{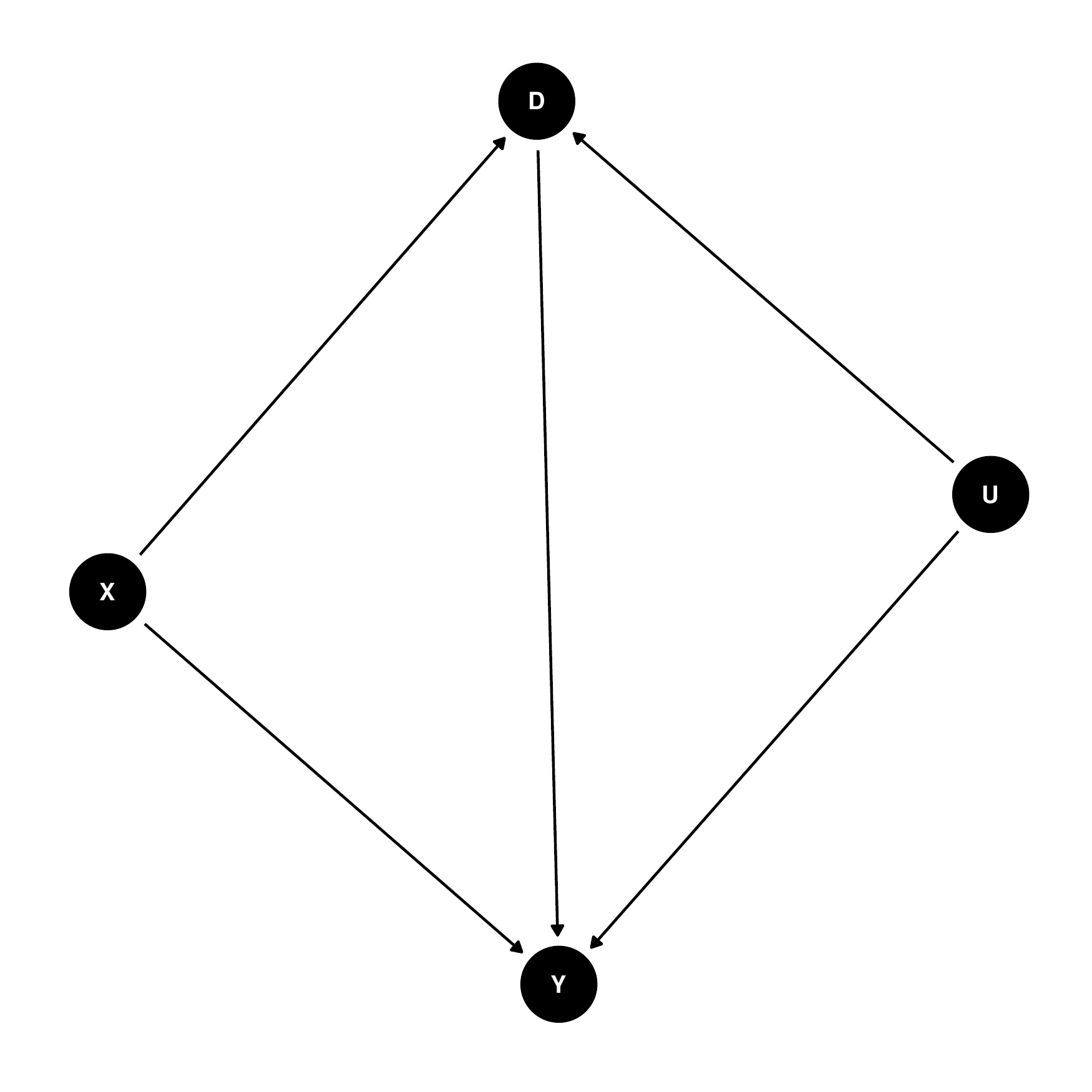}
\caption{$X$ affects both $D$ and $Y$}
\label{fig:dag3}
\end{subfigure}
\caption{Models with a covariate.}
\label{fig:dagx}
\end{figure}

In particular, we are interested in the situations where $X\to Y$ and $Y \leftarrow X \to D$ (see Figure \ref{fig:dag2} and Figure \ref{fig:dag3} respectively). In the first case, $X$ does not further interfere with the identification of the direct effect of $D$ on $Y$ (besides the confounding effect of $U$). While in the second case, further confounding effects are avoided by controlling for $X$.

\begin{figure}
\centering
\begin{subfigure}{0.49\textwidth}
\centering
\includegraphics[width = \textwidth]{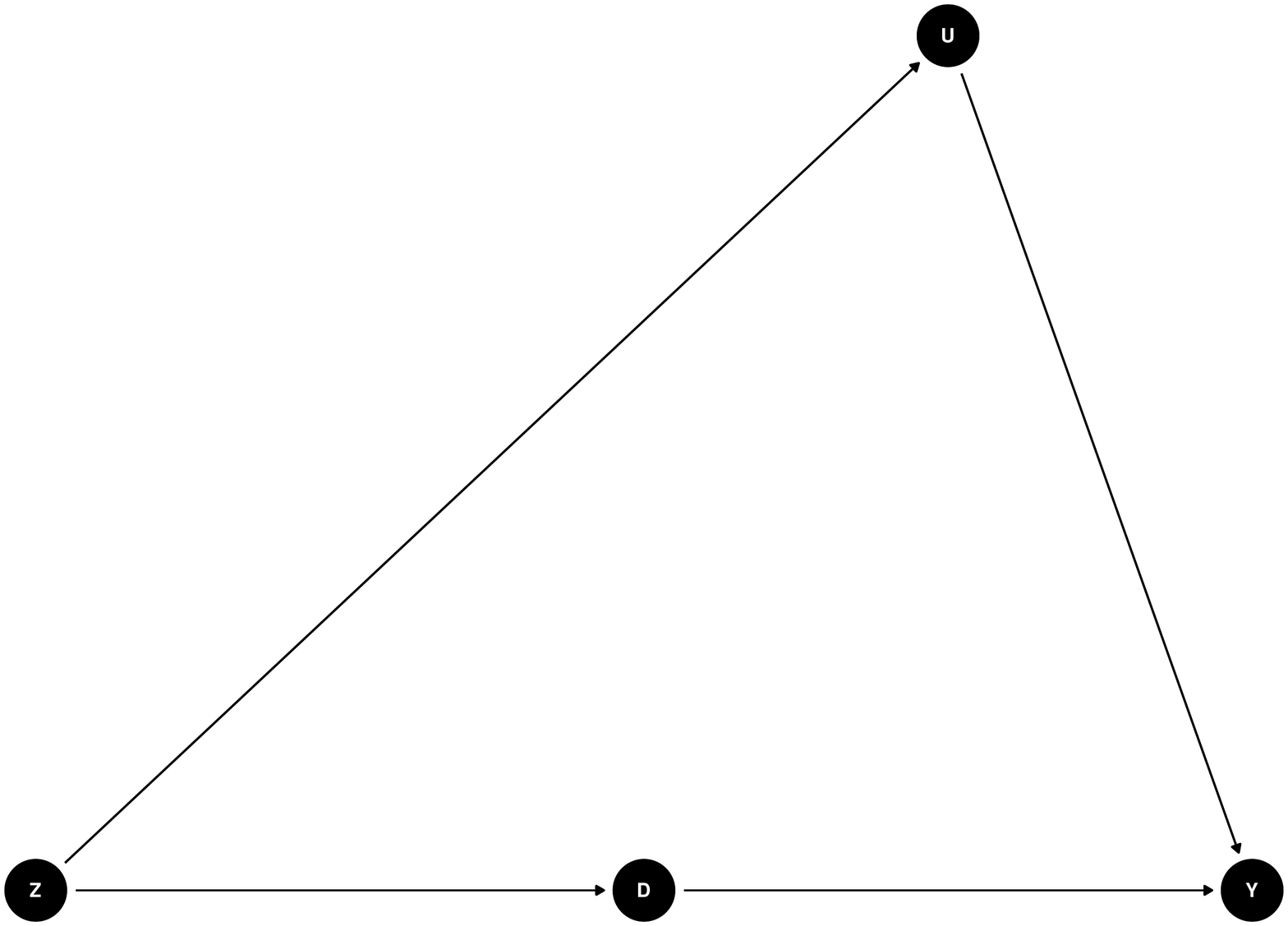}
\caption{$Z$ affects $Y$ through $U$.}
\label{fig:dag-iv1}
\end{subfigure}
\begin{subfigure}{0.49\textwidth}
\centering
\includegraphics[width = \textwidth]{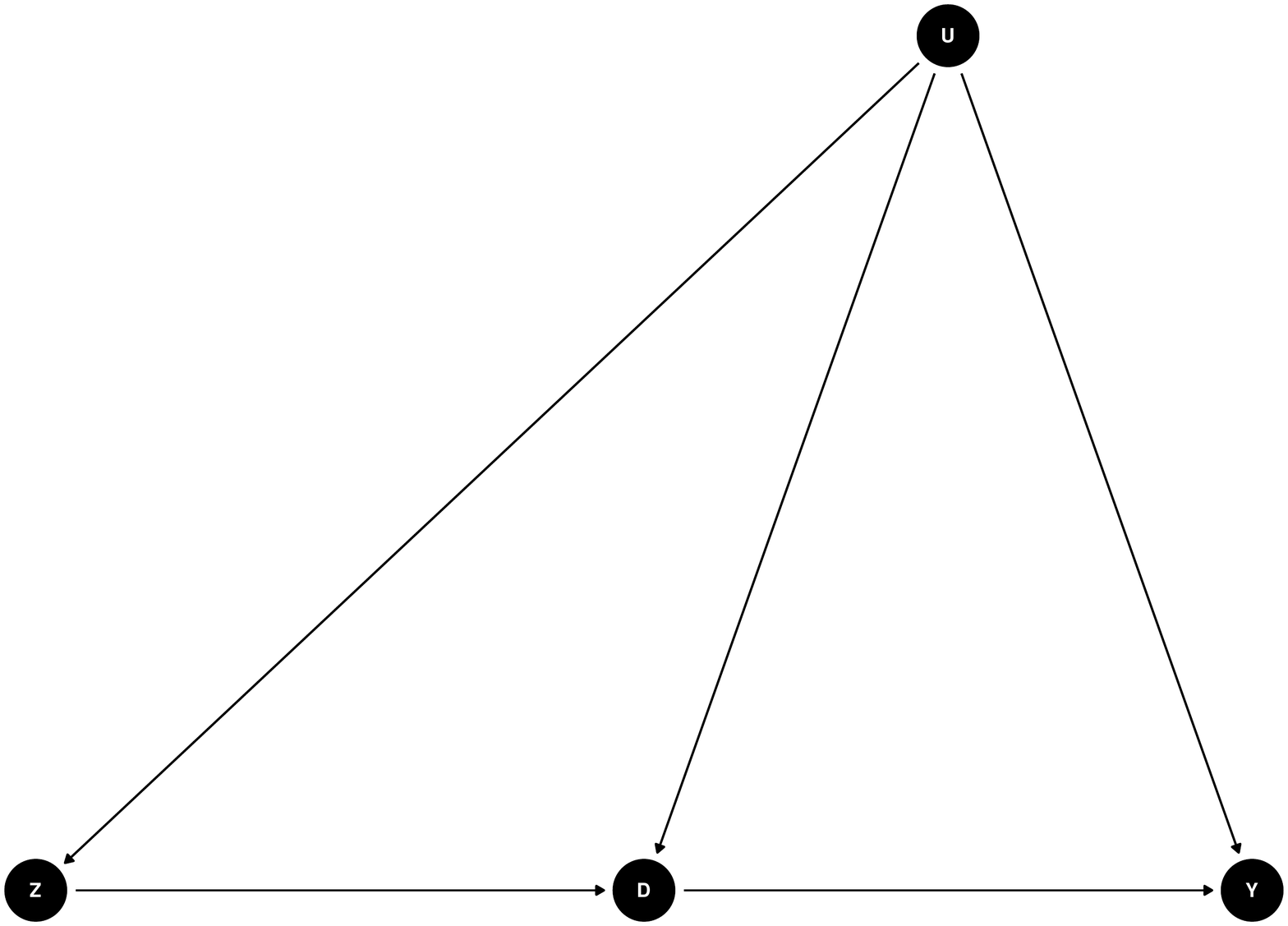}
\caption{$U$ affects both $Z$ and  $Y$.}
\label{fig:dag-iv2}
\end{subfigure}
\caption{Models with an IV.}
\label{fig:dagx}
\end{figure}

When using instrumental variables, it can happen that the instrument $Z$ affects the outcome $Y$  through the variable $U$ ($Z\to U \to Y$), as in Figure \ref{fig:dag-iv1}. In this DGP, $Z$ is not a valid IV for $D$. An alternative situation in which the instrument is not valid, arises when the variable $U$ has an effect on $Z$ opening a confounding path ($Z\leftarrow U \to Y$) as shown in Figure \ref{fig:dag-iv2}.

\section{Monte Carlo experiments}\label{sec:sim}
In this Section, we provide numerical evidence to support our theoretical results.\footnote{As the panel data model presented in the main text is equivalent to a simple linear model once we take the first difference, we do not consider such a model in our Monte Carlo experiments.} For ease of reference one may interpret $Y$ as $wage$, $D$ as $education$ (or $college\;enrollment$ if $D$ is binary) and $U$ as $ability$, typically an unobserved characteristic: we are interested in studying the effect of education on wages, two observed variables, and we assume that unobserved ability may have an effect on wages.   We consider two cases, the first where the treatment variable $D$ is continuous and the second where it is discrete. In addition, we study the properties of the two-stage least squares (2SLS) estimator when the chosen instrument is not valid. In all the examples we will notice that the covariance between the disturbances and either the regressors or the instruments is zero, yet the estimated parameter has no causal interpretation. In all examples, the sample size is set to $n=1000$ and the number of repetitions is 10000.
\subsection{A DGP with a continuous treatment}\label{sec:dgpc}
\begin{figure}
\centering
\begin{subfigure}{0.49\textwidth}
\centering
\includegraphics[width = \textwidth]{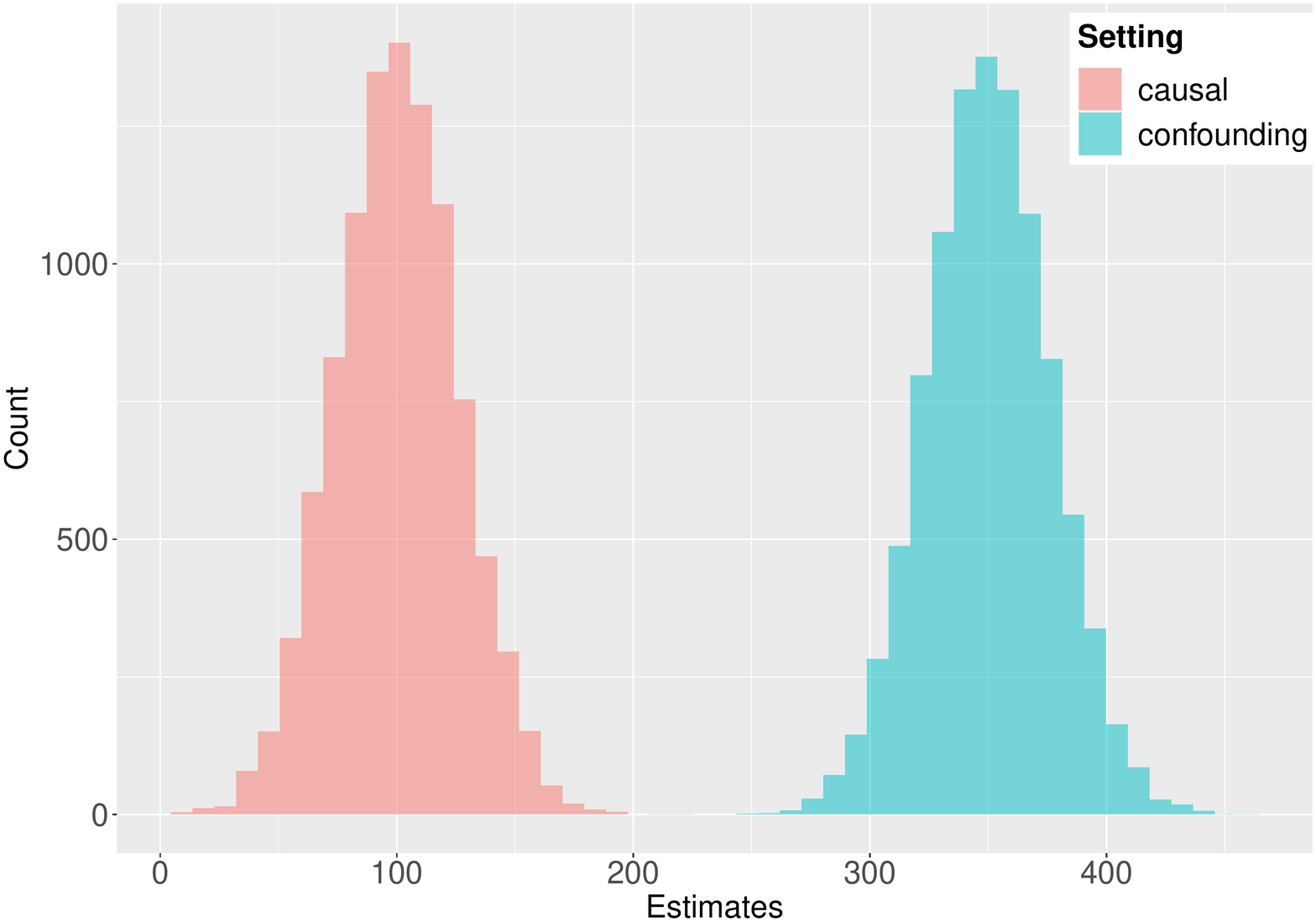}
\caption{OLS estimates of $\tau$ and $\lambda$.}
\label{fig:olscontestimates}
\end{subfigure}
\begin{subfigure}{0.49\textwidth}
\centering
\includegraphics[width = \textwidth]{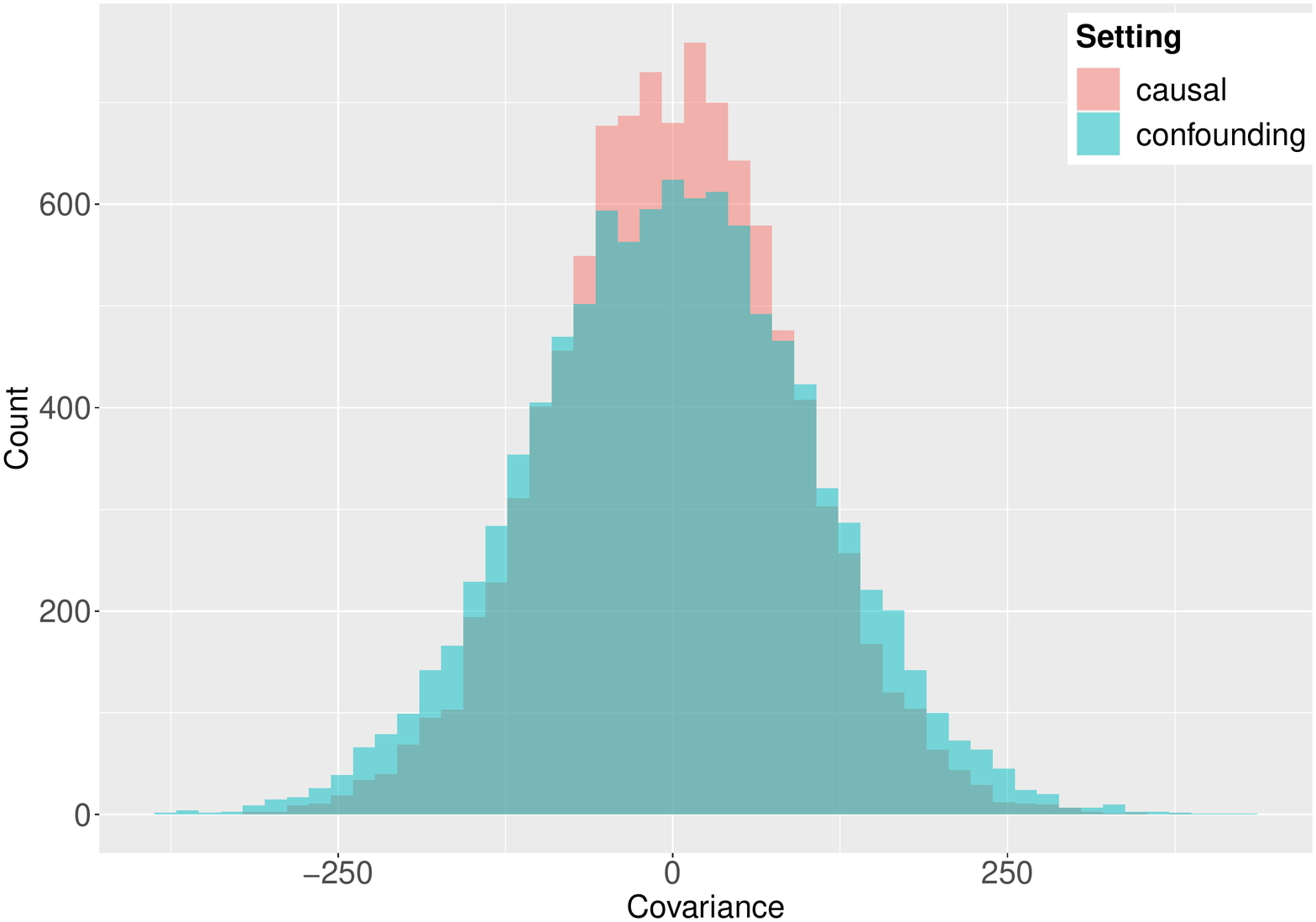}
\caption{Covariance between $v_i$ and $D_i$ and $\varepsilon_i$ and $D_i$.}
\label{fig:olscontcovariance}
\end{subfigure}
\caption{Distribution of estimates and covariances for the linear model with continuous treatment.}
\label{fig:olscont}
\end{figure}

Let us assume the data are produced via the following DGP:
\begin{align}
Y_i = \alpha + \tau D_i + \beta U_i + v_i
\end{align}
with $\alpha = 5000$,  $\tau = 100$, $\beta = 1000$, $v_i \sim (\chi^2_1 - 1) \cdot 1000$, and $(D_i,U_i)' \sim N(\mu,\Sigma)$, with $\mu = (12,0)'$ and 
\begin{align*}
\Sigma = \left(\begin{array}
{rrr}
4 & 1  \\
1 & 1 
\end{array}\right).
\end{align*}
The causal linear conditional expectation function is then $E[Y_i | D_i,U_i] = 5000 + 100 D_i + 1000 U_i$. The non-causal, but still linear, conditional expectation function omitting ability is
\begin{align*}
E[Y_i | D_i] = \left(5000 + 1000 \times \left(0 -  \frac{1}{2}  \frac{1}{2} 12\right)\right ) +\left (100 + 1000 \frac{1}{2} \frac{1}{2}\right) \times D_i = 2000 + 350 \times D_i.
\end{align*}
In this simulation we check the distribution of the estimated coefficients with and without controlling for $U_i$ (Figure \ref{fig:olscontestimates}). We also look at $Cov[v_i,D_i]$ and $Cov[\varepsilon_i,D_i]$ (Figure \ref{fig:olscontestimates}).
We observe that without controlling for ability, the OLS estimator overestimates the returns to education substantially and centers around 350 instead of the true value of 100. However, the covariance of the error term and $D$ center around zero for both the {\it correctly} and the {\it incorrectly} specified regression. This illustrates that $E[\varepsilon_i|D_i] = E[\varepsilon_i] = Cov[\varepsilon_i,D_i] = 0$ is not sufficient to estimate the causal effect correctly.

\begin{figure}
\centering
\begin{subfigure}{0.49\textwidth}
\centering
\includegraphics[width = \textwidth]{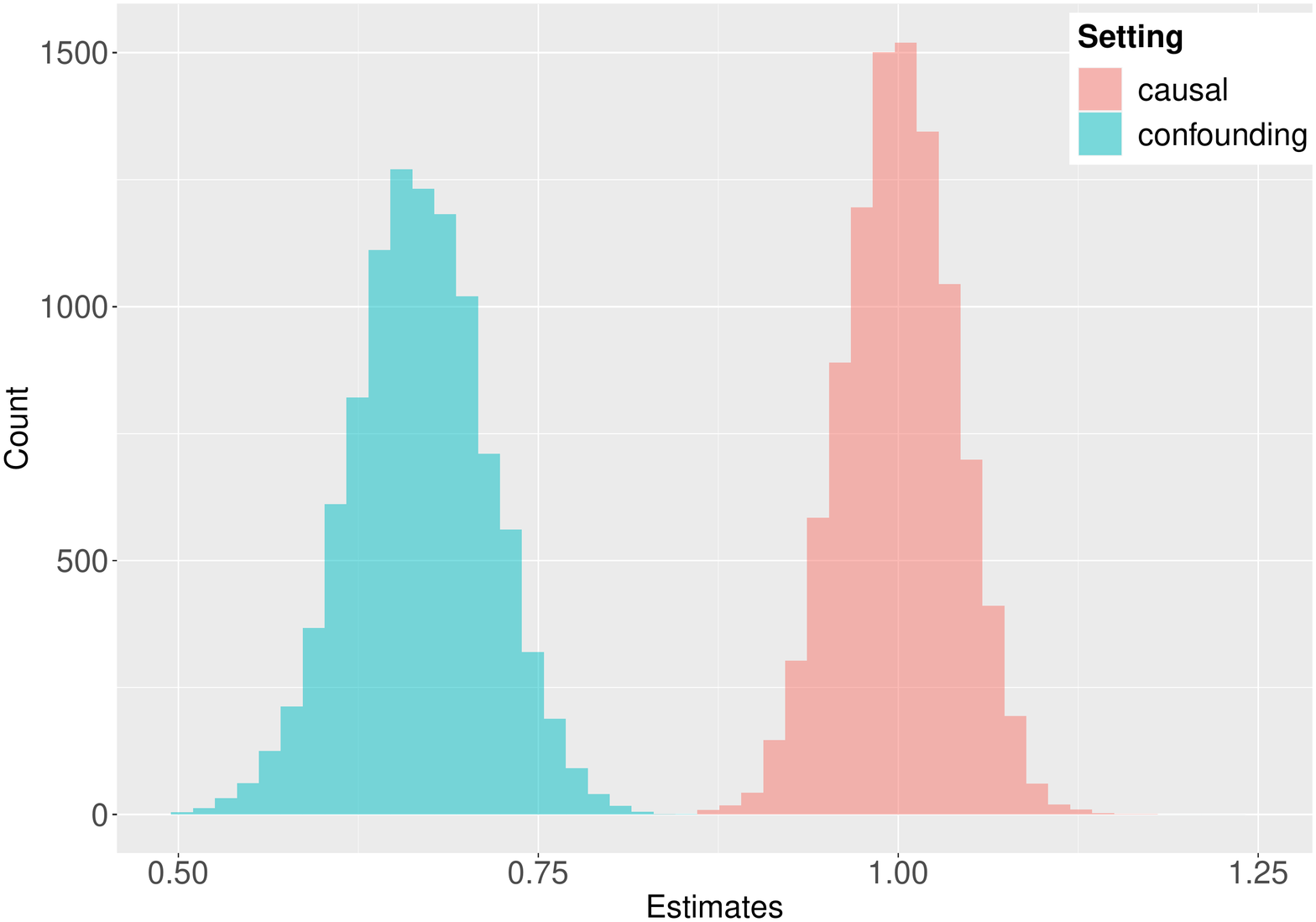}
\caption{OLS estimates of $\tau$ and $\lambda$.}
\label{fig:olsxestimates}
\end{subfigure}
\begin{subfigure}{0.49\textwidth}
\centering
\includegraphics[width = \textwidth]{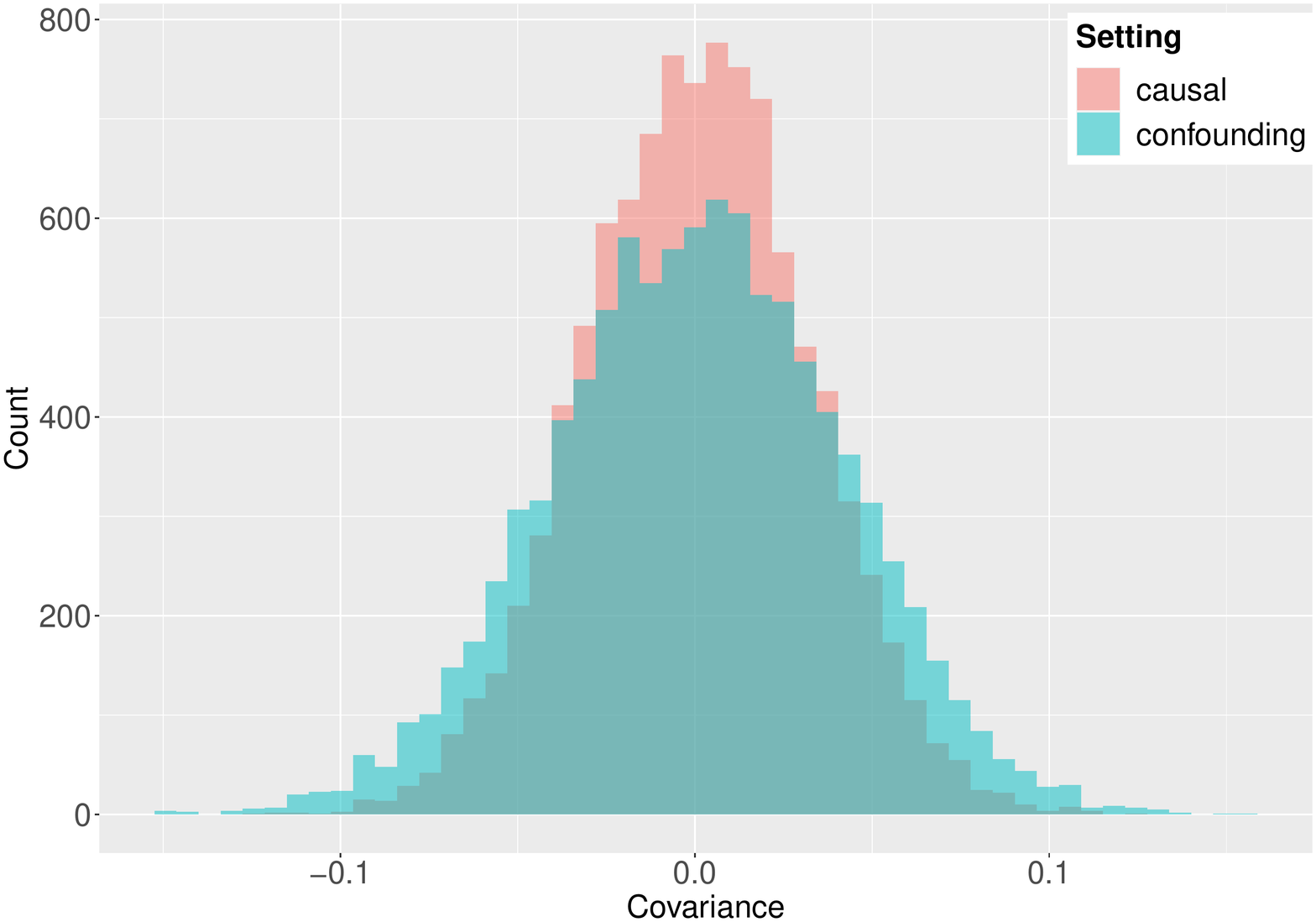}
\caption{Covariance between $v_i$ and $D_i$ and $\varepsilon_i$ and $D_i$.}
\label{fig:olsxcovariance}
\end{subfigure}
\caption{Distribution of estimates and covariances for the linear model with continuous treatment and exogenous controls.}
\label{fig:olsx}
\end{figure}

%We shall see that even if we include additional control variables the problem also occurs
We shall see that this problem also occurs in a setting with control variables if there is remaining unobserved confounding (see Figure \ref{fig:dag2} and Figure \ref{fig:dag3}).
%It is often the case that practitioners include additional controls as a means to avoid confounding (see Figure \ref{fig:dag2} and Figure \ref{fig:dag3}). Even in this case we shall see that without proper causal assumptions the identification of the causal effect is impossible \citep[see][]{angrist2009mostly,cfp2022}.  
For illustration, let us define the DGP as 
\begin{align}
Y_i= \alpha+\tau D_i + \phi X_i + \beta U_i +v_i
\end{align}
where the $3\times 1$ vector $(D_i,X_i, U_i)'$ is jointly normally distributed with zero mean and covariance matrix $\Sigma$; the diagonal entries of $\Sigma$ are equal to 1 and the off-diagonal entries are set to $\rho=1/2$. Moreover, $\alpha=0$ and $\tau=\delta=\beta=1$. In turn we estimate 
\begin{align}
Y_i= \gamma +\lambda D_i + \varphi X_i + \varepsilon_i.
\end{align}
As in the previous case, we notice that the covariances are centered around zero (Figure \ref{fig:olsxcovariance}), but estimating $\lambda$ by OLS would not provide unbiased estimates of $\tau$, the causal effect of interest (Figure \ref{fig:olsxestimates}).

\subsection{A DGP with a discrete treatment}\label{sec:dgpd}
Consider the following setting with discrete treatment. Let two covariates be drawn from a multivariate standard normal distribution $(X_i,U_i)' \sim N(\mu,\Sigma)$ with $\mu = (0,0)'$ and 
\begin{align*}
\Sigma = \left[\begin{array}
{rrr}1 & \rho  \\
\rho  & 1
\end{array}\right]
\end{align*}
where $\rho$ is the correlation coefficient between $X_{i}$ and $U_{i}$. The binary treatment is then created as $D_i = 1[X_{i}>c]$. 

\begin{figure}
\centering
\begin{subfigure}{0.49\textwidth}
\centering
\includegraphics[width = \textwidth]{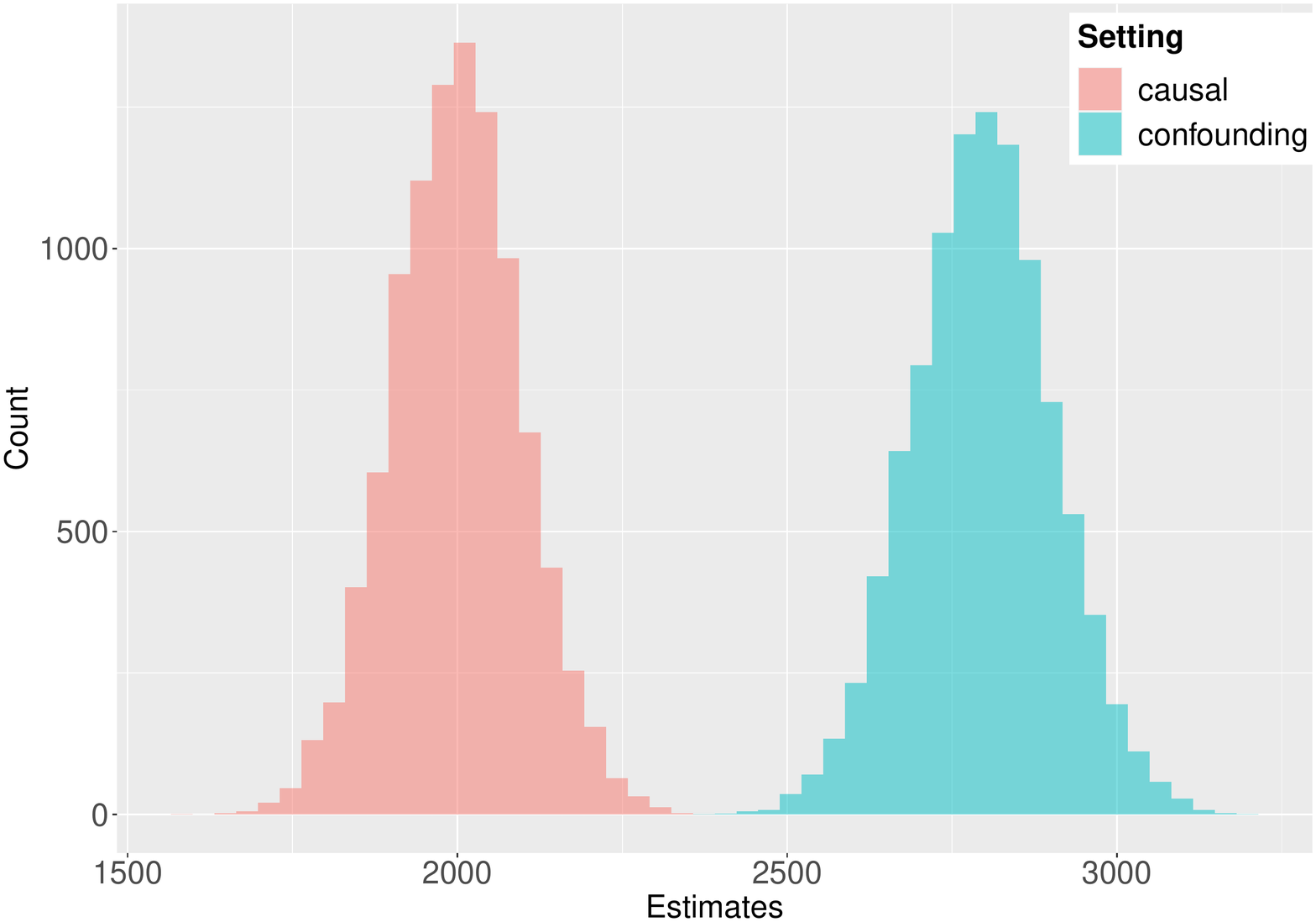}
\caption{OLS estimates of $\tau$ and $\lambda_1$.}
\label{fig:olsdestimates}
\end{subfigure}
\begin{subfigure}{0.49\textwidth}
\centering
\includegraphics[width = \textwidth]{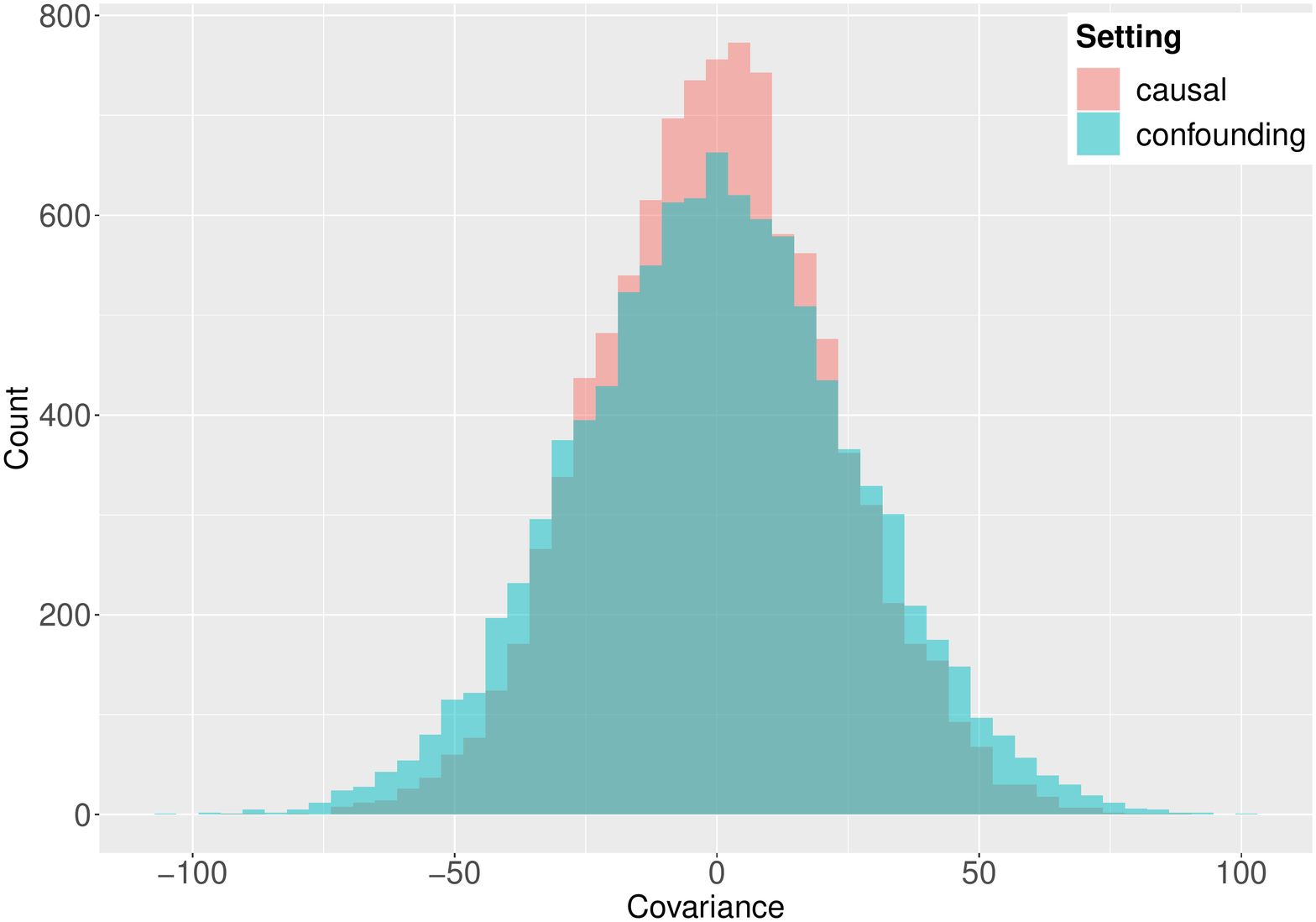}
\caption{Covariance between $v_i$ and $D_i$ and $\varepsilon_i$ and $D_i$.}
\label{fig:olsdcovariance}
\end{subfigure}
\caption{Distribution of estimates and covariances for the linear model with binary treatment.}
\label{fig:olsd}
\end{figure}

\begin{figure}
\centering
\begin{subfigure}{0.49\textwidth}
\centering
\includegraphics[width = \textwidth]{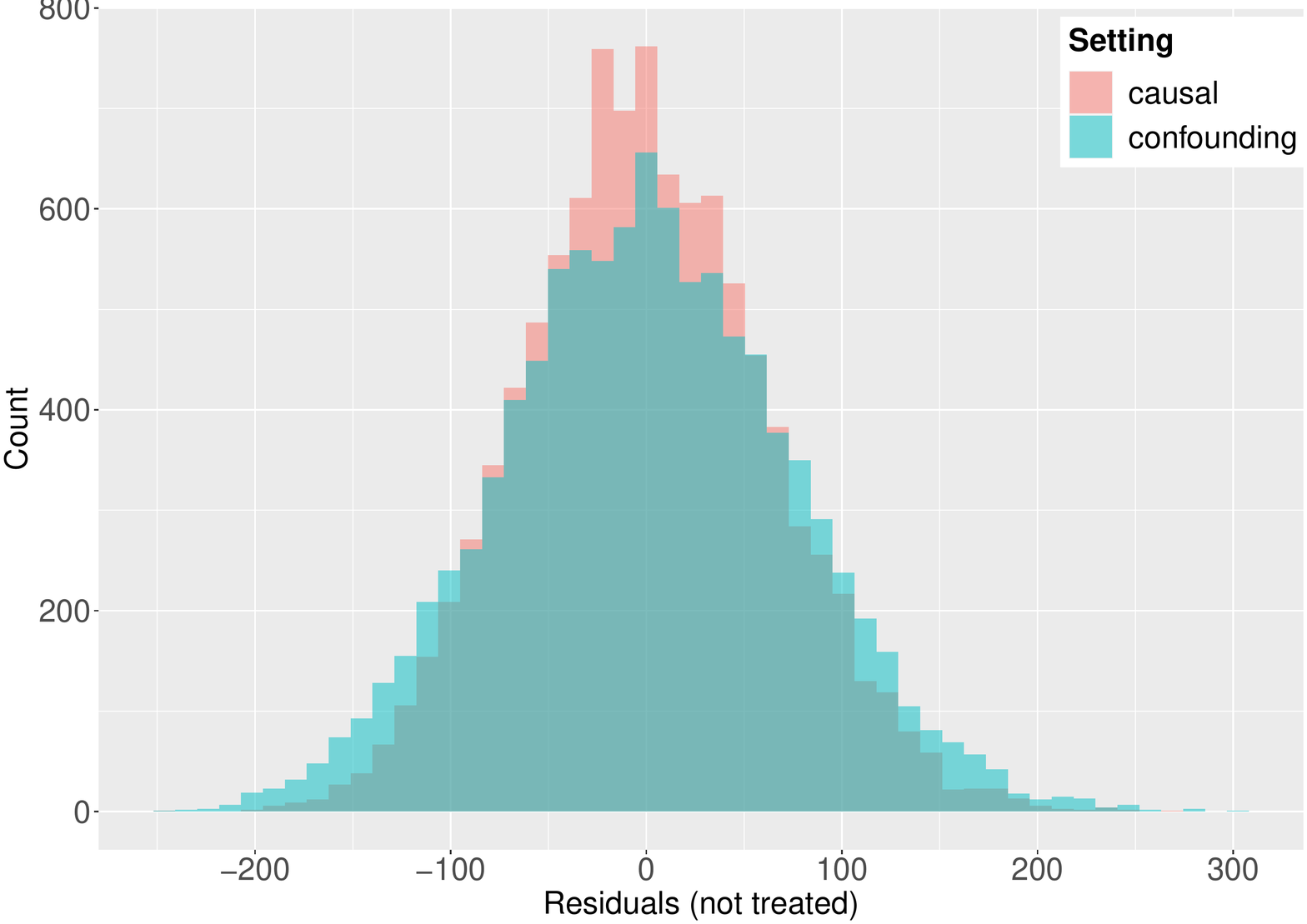}
\caption{Residuals for $D_i=0$.}
\label{fig:olsd0}
\end{subfigure}
\begin{subfigure}{0.49\textwidth}
\centering
\includegraphics[width = \textwidth]{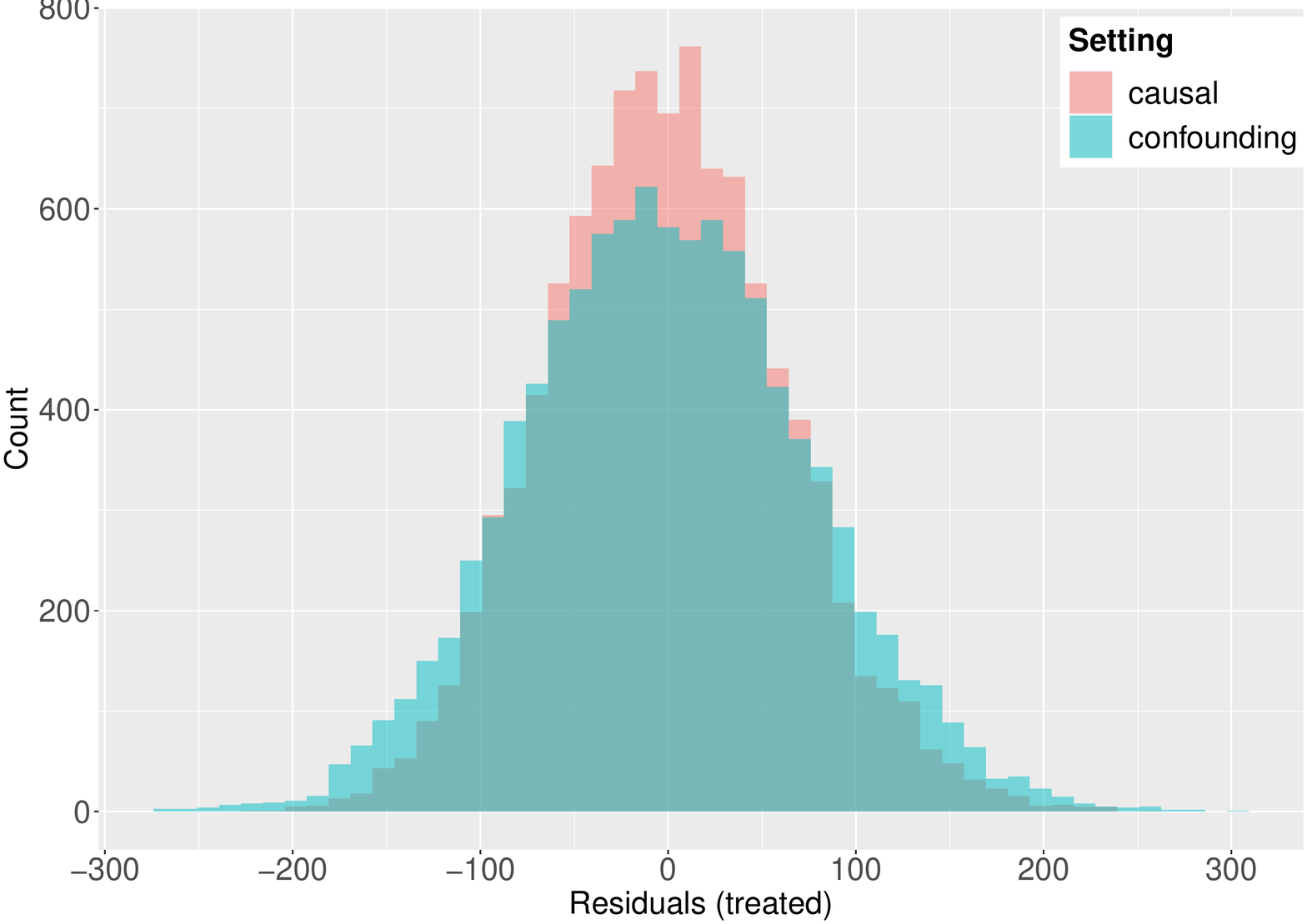}
\caption{Residuals for $D_i=1$.}
\label{fig:olsd1}
\end{subfigure}
\caption{Distribution of residuals for untreated and treated units.}
\label{fig:olsd01}
\end{figure}

Now assume that the DGP is:
\begin{align*}
Y_i = \alpha + \tau  D_i + \beta  U_i + v_i
\end{align*}
with $\alpha = 5000$,  $\tau = 2000$, $\beta = 1000$, $u \sim (\chi^2_1 - 1) \cdot 1000$, $\rho = 1/2$, and $c=0$. The causal linear conditional expectation function is then $E[Y_i | X_i,U_i] = 5000 + 2000 D_i + 1000 U_i$. The non-causal, but still linear, conditional expectation function omitting $U_i$ is
\begin{align*}
E[Y_i | D_i] = \left(5000 - 1000 \frac{1}{2}  \frac{\phi(0)}{\Phi(0)} \right) + \left(2000 + 1000 \frac{1}{2} 2 \frac{\phi(0)}{\Phi(0)}\right) \times D_i \approx 4601 + 2798 \times D_i.
\end{align*}
We run a simulation study by estimating 
\begin{align*}
Y_i=\gamma +\lambda D_i+\varepsilon_i
\end{align*}
and we check the distribution of the estimated coefficients with and without controlling for ability as well as $Cov(u_i,D_i)$, $Cov(\varepsilon_i,D_i)$, $E(u_i|D_i = 0)$, $E(\varepsilon_i|D_i = 0)$, $E(u_i|D_i = 1)$, and $E(\varepsilon_i|D_i = 1)$. Also in this case the results (Figure \ref{fig:olsd}) are in line with those found for the continuous treatment case. Furthermore, the average disturbances for $D_i=0$ and $D_i=1$ remain centered around zero both for the estimated model and the actual (causal) outcome model (Figure \ref{fig:olsd01}).

\subsection{A DGP with instrumental variables}\label{sec:dgpe}
In this section, we consider the case where a researcher mistakenly considers an invalid IV, similarly to the case described in Figure \ref{fig:dag-iv1}.\footnote{Notice that a DGP similar to  Figure \ref{fig:dag-iv2} requires a more cumbersome parametrization but yields analogous results, which are available from the authors upon request.  }  We will see that even in this case the causal effect cannot be recovered even if the correlation between the disturbances of the considered model and the used instrument is zero. The data are generated according to the equations
\begin{align}\label{DGP3}
Y_i&=\alpha+\tau D_i+\beta U_i+v_i\\
D_i&=\pi Z_i +  \eta_i\\
U_i&=\mu+\delta Z_i + \xi_i
\end{align}
\begin{figure}
\centering
\begin{subfigure}{0.50\textwidth}
\centering
\includegraphics[width = \textwidth]{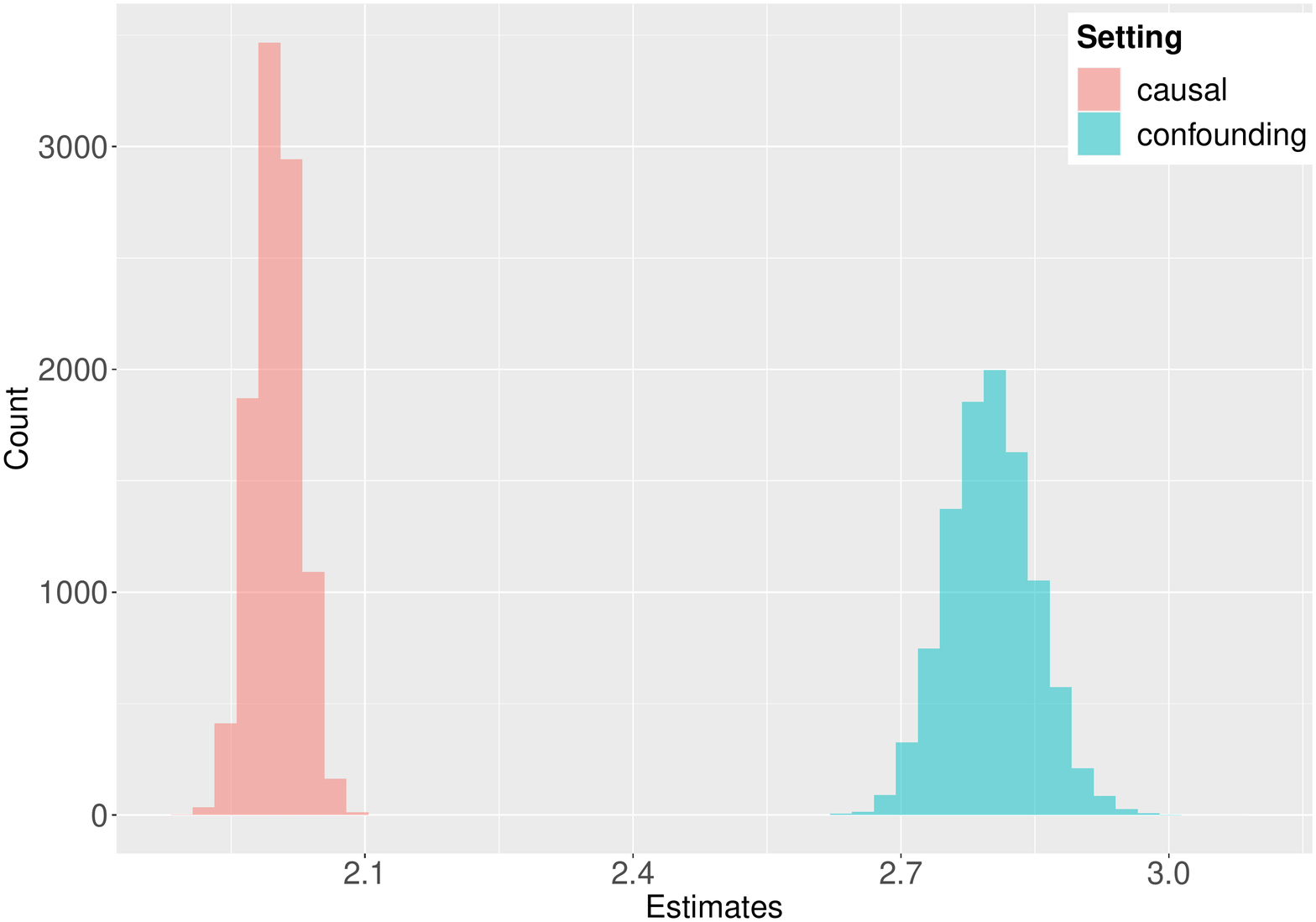}
\caption{OLS estimates of $\tau$ and 2SLS estimates of $\lambda$.}
\label{fig:ivestimates}
\end{subfigure}
\begin{subfigure}{0.49\textwidth}
\centering
\includegraphics[width = \textwidth]{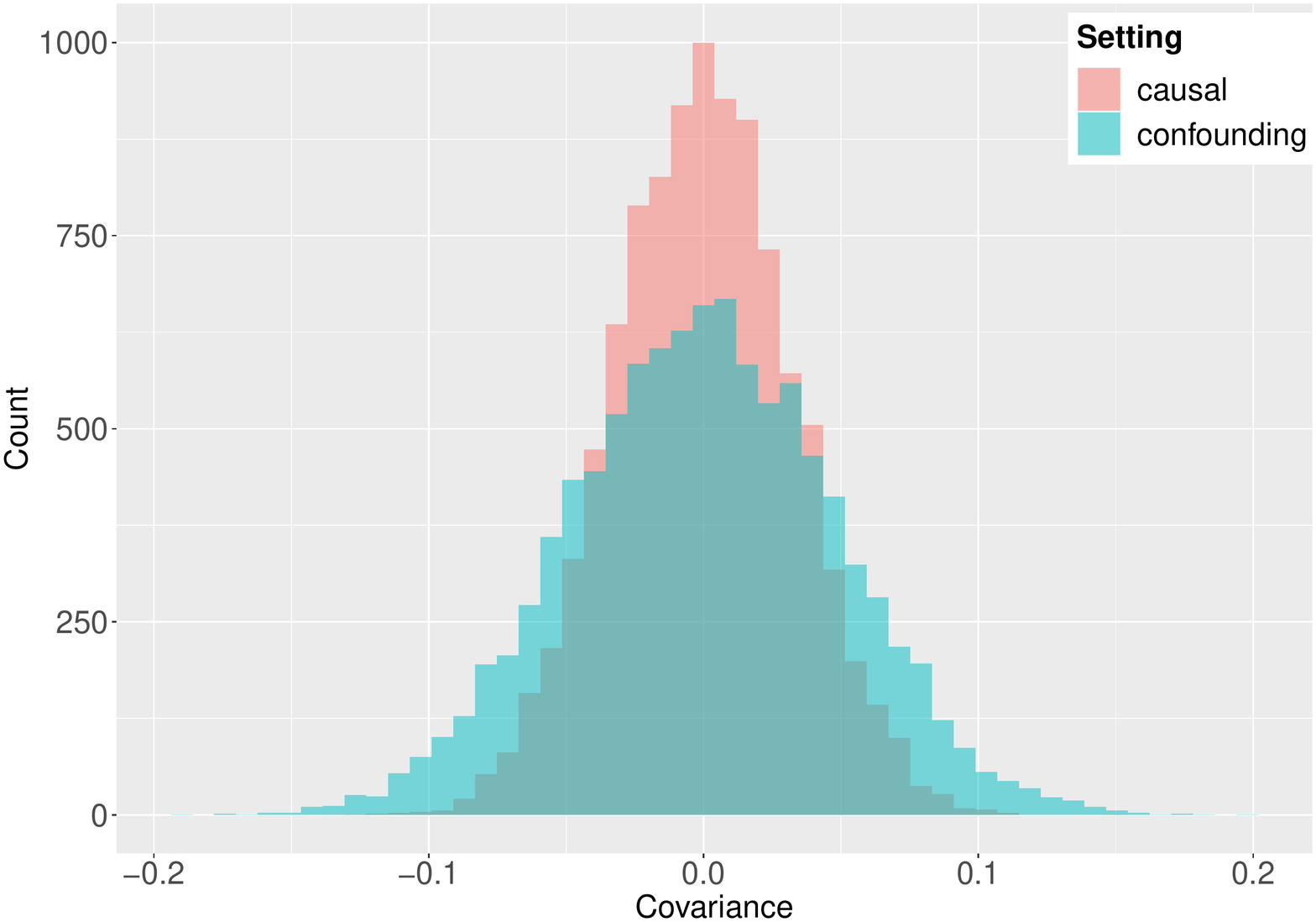}
\caption{Covariance between $v_i$ and $Z_i$ and $\varepsilon_i$ and $Z_i$.}
\label{fig:ivcovariance}
\end{subfigure}
\caption{Distribution of  estimates and covariances for the linear model with instrumental variables.}
\label{fig:iv}
\end{figure}

and we are interested in recovering the effect $D \to Y$ represented by $\tau$. The variables $U_i$, $v_i$, $\eta_i$ and $\xi_i$ are sampled independently from a standard normal distribution, while the parameters are chosen as $\alpha=\mu=0$,$\tau=\gamma=2$, $\beta=.8$, $\pi=\delta=1$. We estimate the model 
\begin{align*}
Y_i=\gamma + \lambda D_i+\varepsilon_i
\end{align*}
by 2SLS with $Z_i$ being the instrumental variable. 
The causal linear conditional expectation function is then $E[Y_i | D_i,U_i] =  2 \times D_i + .8 \times U_i$. The non-causal, but still linear, conditional expectation function omitting $U_i$ is
\begin{align*}
E[Y_i | D_i] = (0+0\times .8)+\left(2 + \frac{.8\times 1}{1}\right)\times D_i =  2.8 \times D_i.
\end{align*}

The results illustrated in Figure \ref{fig:iv} are in line with the other examples provided so far: we find that the correlation between the instrument $Z_i$ and the disturbances is centered around zero (Figure \ref{fig:ivcovariance}) but we are unable to recover the true causal effect (Figure \ref{fig:ivestimates}).

\end{appendices}
\end{document}